\documentclass[article]{aa}  

\usepackage{graphicx}
\usepackage{subfigure}
\usepackage{txfonts}
\usepackage[dvipsnames]{xcolor}
\usepackage{hyperref}
\definecolor{coolblack}{rgb}{0.0, 0.18, 0.39}
\definecolor{darkblue}{rgb}{0.0, 0.0, 0.55}
\definecolor{mediumred-violet}{rgb}{0.78, 0.08, 0.52}
\hypersetup{
    colorlinks=true, 
    linkcolor=darkblue,
    filecolor=magenta,      
    urlcolor=darkblue,
    citecolor=darkblue 
}
\usepackage[dvipsnames]{xcolor}

\usepackage{dblfloatfix}
\usepackage[separate-uncertainty=true]{siunitx}
\newlength\figureheight
\newlength\figurewidth
\usepackage[switch]{lineno}
\DeclareSIUnit\year{yr}

\newcommand{\C}{3C\,84}
\newcommand{\new}[1]{\textcolor{black}{#1}}

\setcitestyle{citesep={,}}
\begin{document}

   \title{Spectral index evolution of the limb-brightened jet in 3C\,84}

   \author{
   L.~C. Debbrecht \inst{\ref{mpifr}},  
   G.~F. Paraschos\inst{\ref{Finca},\ref{Mh},\ref{mpifr}}, 
   E. Ros\inst{\ref{mpifr}},
   T.~P. Krichbaum\inst{\ref{mpifr}},
   U. Bach\inst{\ref{mpifr}}, 
   M.~A. Gurwell\inst{\ref{CfA}}, 
   J.~A. Hodgson\inst{\ref{Seoul}}, 
   M. Janssen\inst{\ref{IMAPP},\ref{mpifr}},  
   J.-Y. Kim\inst{\ref{UNIST}}, 
   M.~M. Lisakov\inst{\ref{IFIS}}, 
   N.~R. MacDonald\inst{\ref{UMis},\ref{mpifr}}, 
   D.~G. Nair\inst{\ref{Chile},\ref{mpifr}}, 
   J.~Oh\inst{\ref{JIVE}}, 
   J.~A. Zensus\inst{\ref{mpifr}}
          }
    \authorrunning{L.~C. Debbrecht et al.}
   \institute{
          Max-Planck-Institut f\"ur Radioastronomie, Auf dem H\"ugel 69, D-53121 Bonn, Germany\label{mpifr} \\
          \email{ldebbrecht@mpifr-bonn.mpg.de}\and
          Finnish Centre for Astronomy with ESO, University of Turku, 20014 Turku, Finland\label{Finca}\and
          Aalto University Metsähovi Radio Observatory, Metsähovintie 114, FI-02540 Kylmälä, Finland\label{Mh}\and
          Center for Astrophysics | Harvard \& Smithsonian, 60 Garden Street, Cambridge, Massachusetts, 02138 USA\label{CfA}\and
          Dept. of Physics \& Astronomy, Sejong University, Guangjin-gu, Seoul 05006, Republic of Korea\label{Seoul}\and
          Department of Astrophysics, Institute for Mathematics, Astrophysics and Particle Physics (IMAPP), Radboud University, P.O. Box 9010, 6500 GL Nijmegen, The Netherlands\label{IMAPP}\and
          Ulsan National Institute of Science and Technology, 50 UNIST-gil, Eonyang-eup, Ulju-gun, Ulsan 44919, Republic of Korea\label{UNIST}\and
          Instituto de Física, Pontificia Universidad Católica de Valparaíso, Casilla 4059, Valparaíso, Chile\label{IFIS}\and
          Department of Physics and Astronomy, University of Mississippi, University, Mississippi 38677, USA\label{UMis}\and
          Astronomy Department, Universidad de Concepción, Casilla 160-C, Concepción, Chile\label{Chile}\and
          Joint Institute for VLBI ERIC (JIVE), Oude Hoogeveensedijk 4, 7991 PD Dwingeloo, The Netherlands\label{JIVE}
             }
   \date{Received XX; accepted YY}

  \abstract{
  Relativistic jets launched by active galactic nuclei are fundamental for understanding the physics of accreting supermassive black holes and their immediate environment, yet the mechanisms driving jet launching remain uncertain. 
  In this study, we investigate the sub-parsec jet of 3C\,84 using multi-epoch, multi-frequency, very long baseline interferometry (VLBI) observations \new{with the European VLBI Network and the Very Long Baseline Array} at 22 and 43\,GHz. 
  We analyse the evolution of the spectral index gradient in the core region to relate the observed structure to physical interpretations and to discriminate between competing jet launching models. 
   Furthermore, we examine the impact of the ambient medium and magnetic field configuration on jet morphology and dynamics over time, and explore their connection to a coinciding $\gamma$-ray flare.
   Our spectral analysis reveals significant changes across three epochs, indicating dynamic activity between filamentary structures on sub-parsec scales, evolving magnetic fields, and a complex interaction with the surrounding medium, all of which shape the innermost jet and may influence its high-energy emission. 
   }

   \maketitle
   \nolinenumbers
\section{Introduction}
Relativistic jets in active galactic nuclei (AGN) are highly collimated and energetic outflows of plasma, emanating from the vicinity of the central engine. 
Until today, the exact launching process of relativistic jets remains a subject of investigation. 
The launching mechanism postulated by \citeauthor{BZ} (\citeyear{BZ}; BZ process) suggests that rotational energy is directly extracted from the central black hole, launching a jet close to the central engine.
On the contrary, \citeauthor{BP} (\citeyear{BP}; BP process) predicts that rotational energy is extracted from the accretion disc, powering a jet which has a wider and stratified jet base. 
\new{However, these two mechanisms are not mutually exclusive and may coexist, potentially contributing jointly to the observed jet structure.} 
AGN jets have been widely tested for these mechanisms, in which a more collimated jet indicates the presence of the BZ process~\citep[see][for a review]{Walker2018, Lu2023}, while a wider jet  would favour the BP mechanism~\citep{Giovannini2018, GP2024EHT}.
Since these two launching models result from a substantially different magnetic field configuration and jet geometry, they can be distinguished by studying the innermost jet region of nearby AGN. 
Very long baseline interferometry (VLBI) observations at centimetre and millimetre wavelengths provide high angular resolution, that is essential for probing the innermost regions of AGN jets, where jet collimation takes place~\citep[see][for more details]{Boccardi2021}. \\
\C\, is a compact radio source in the central galaxy NGC\,1275 of the Perseus cluster and due to its proximity at redshift $z=0.018$~\citep{Strauss92}, its brightness, and jet structure, it is a prime target for testing jet launching scenarios, as well as studying the innermost sub-parsec AGN structure~\citep{Walker2000, Kam2024}. 
Previous VLBI studies reveal a complex jet morphology in the sub-mas region:
\cite{GP2022} observed a time-varying spectral index gradient in the innermost jet (2011–2020), attributed to structural variability, while \citet{Park2024} reported transverse limb-brightening gradients with inverted edge spectra, indicating complex ambient-medium interactions. 
\cite{Nagai2014}, \cite{Giovannini2018}, \cite{Kim2019}, and \cite{GP2024} found evidence for a limb-brightened structure in total intensity and in linear polarisation, whereby the edge appears brighter than the spine, consistent with a stratified jet, a velocity or magnetic field gradient, or a combination of both~\citep[for a detailed review see][]{Blandford2019}. 
\new{Those filamentary structures starting from the core region, are collimated and elongated patterns of enhanced emission, extending into the jet further downstream~\citep{Nagai2014, Giovannini2018, Punsly2021}.}
\cite{GP2025} investigated these filaments close to the central engine, using cm-VLBI observations and found, that their morphology may arise from Kelvin-Helmholtz (K-H) instabilities, driven by interactions between the jet plasma and the ambient medium~\citep{Fuentes2023, Nikonov2023}.  
Additionally, \C\, shows strong $\gamma$-ray variability \citep{Nagai2012, Hodgson2021}.
Previously, the ejection of new components in \C\, have been linked to subsequent $\gamma$-ray flares~\citep{GP2022, GP2023, Hodgson2021}, with a delayed flare peak, however, the spatial origin of $\gamma$ rays in this source is thought to be diverse, in both the VLBI core and the downstream jet, where the jet shows signatures of strong interaction with the interstellar medium. 
\\
In this study we aim to investigate a possible connection between the observed limb-brightening and the variability in the $\gamma$-ray regime by probing the innermost structures with multi-epoch, multi-frequency VLBI observations of \C.
In particular, we examine whether the spectral index distribution within the jet of \C\, also exhibits limb-brightened features similar to the electric vector position angle (EVPA) patterns observed in the jet's linear polarisation strength \citep[as seen in][]{GP2024}. 
Moreover, we aim to image the sub-parsec scale region, to map the spectral index gradient, \new{which provides insights into the distinction of different jet launching models}~\citep{GP2022proceedings}. 
Leveraging the increased sensitivity of our latest VLBI datasets, we examine temporal changes in the spectral index with high precision to identify signatures of synchrotron emission, absorption, and particle acceleration, \new{hereby providing possible explanations for the jet’s morphology} and its interaction with the surrounding medium. 
We further compare radio and $\gamma$-ray light-curves to link changes in high-energy emission to structural and spectral variations, providing insights into the connection between flaring events and the jet launching region.
\\
This paper is structured as follows: In Sect.~\ref{Sec:ObsDataRed} we provide information about the observations and data reduction. 
Sect.~\ref{Sec:MethodsResults} outlines the methods and presents the results. 
In Sect.~\ref{Sec:Discussion} we discuss our results and in Sect.~\ref{Sec:Conclusion} we draw  conclusions of our analysis. 
Throughout this paper we assume a $\Lambda$ cold dark matter cosmology with $H_0 = 67.8\,\mathrm{km\,s}^{-1}\,\mathrm{Mpc}^{-1}$, $\Omega_\Lambda = 0.692$, and $\Omega_\mathrm{m} = 0.308$~\citep{PlanckCollab}, so that $1\,\mathrm{mas}$ corresponds to $0.37~\mathrm{pc}$.

\section{Observations and data reduction \label{Sec:ObsDataRed}}
\C\, was observed by the European VLBI Network (EVN) in June 2024 and November 2021 at 22 and 43\,GHz, supplemented by observations by the Very Long Baseline Array (VLBA) and the Global VLBI Alliance (GVA) in 2022 at the same frequencies. 
\new{Due to observational issues affecting the 43\,GHz data from 2021 on long baselines, we utilised data available from the VLBA-BU Blazar Monitoring Program (BEAM-ME and VLBA-BU-BLAZAR\footnote{\url{http://www.bu.edu/blazars/BEAM-ME.html}}). }
The data from observations in 2021, 2022, and 2024 were recorded in eight baseband channels for each frequency, with two-bit quantisation, and a sample rate of 2\,Gbps for all but the Australian Long Baseline Array (LBA) in 2022 (1\,Gbps). 
Finally, the data of 2021, 2022, and 2024 were correlated at the correlator at the Joint Institute for Very Long Baseline Interferometry European
Research Infrastructure Consortium (JIVE) in Dwingeloo, Netherlands. 
We calibrated the data of 2021 and 2024 using the pipeline \texttt{rPICARD}\footnote{Comparative tests demonstrate that \texttt{rPICARD} produces improved data calibration results than the calibration procedure performed with AIPS, as shown in~\cite{Kim2023}.}~\citep{Janssen2019}. 
We used standard VLBI fringe-fitting and calibration procedures, and applied the same steps as explained in~\cite{GP2024}. 
The data of 2022 were used as published by~\cite{Park2024}.
After the calibration, the data of 2021 and 2024 were first averaged over all IFs and then time-averaged in 30-second bins. 
We employed a hybrid imaging technique, iteratively combining \textsc{clean} deconvolution algorithm \citep{CLEANHogbom} in \textsc{difmap}~\citep{Difmap2}, together with self-calibration in phase and amplitude. 
In order to address uncertainties in the absolute amplitudes of the visibilities, we applied a constant systematic non-closing error of 3\% to the data of 2024, as presented in \cite{EHT2019}, to account for low signal-to-noise measurements. 
A summary of all information about the observations and the \textsc{clean}-maps of \C\, are listed in Table~\ref{tab:obs}.\\
Furthermore, we incorporated $\gamma$-ray and radio flux light-curves into our analysis. 
For the $\gamma$-ray emission we utilised publicly available data by the Fermi Large Area Telescope Collaboration\footnote{\url{https://fermi.gsfc.nasa.gov/ssc/data/access/lat/LightCurveRepository/}}~\citep[Fermi-LAT; for a detailed description see][]{Atwood2009,Abdollahi2023FERMI} repository and adopted a monthly cadence. 
The radio flux measurements at 1.3\,mm were obtained from the publicly available data
of \C\, by the Submillimeter Array ~\citep[SMA\footnote{\url{http://sma1.sma.hawaii.edu/callist/callist.html}};][]{SMAHo2004,Gurwell2007}.

\begin{table*}
\centering
\caption{Summary of VLBI observations of \C\, used in this study.}
\begin{tabular}{cccccccccc}
\hline
Epoch & Frequency & Instrument & Beam & Position Angle & $I_\mathrm{peak}$ & $\sigma_\textrm{I}$ & \new{First publication}\\
\text{[yyyy-mm-dd]} & [GHz] &  & [mas] & [$^\circ$] & [Jy/beam] & [mJy/beam] &\\ 
(1)\label{tab:data} & (2)\label{tab:freq} & (3)\label{tab:Intf} & (4)\label{tab:beam} & (5)\label{tab:PA} & (6)\label{tab:Ipeak} & (7)\label{tab:Isigma} & (8) \\ \hline \hline
2021-11-08 & 22 & EVN  & $0.40 \times 0.17$  & $-15.8$ & 1.62 & 0.4 & Ref1 \\
2021-11-10 & 43 & VLBA & $0.32 \times 0.15 $ & $-1.60$ & 2.14 & 1.2 & Ref2\\ \hline
2022-11-09 & 22 & GVA  & $0.45 \times 0.21 $ & $13.4$  & 1.00 & 1.7 & Ref3\\
2022-11-01 & 43 & VLBA & $0.31 \times 0.15 $ & $6.10$  & 2.36 & 4.1 & Ref3 \\ \hline
2024-06-05 & 22 & EVN  & $0.36 \times 0.17 $ & $-2.51$ & 3.54 & 3.0 & This work \\
2024-06-07 & 43 & EVN  & $0.18 \times 0.08 $ & $-7.70$ & 5.94 & 0.7 & This work\\ \hline
\end{tabular}
\tablefoot{(1) Date of observation in year-month-day format. (2) Observing frequency. (3) Interferometer used for observation. (4) The nominal restoring beam sizes of the clean image in mas  (uniform weighting) and (5) the position angle. (6) Total intensity peak of Stokes I in units of Jy per beam (using $(0.35 \times 0.16)$\,mas at a position angle of $18^\circ$). (7) Total intensity RMS level. \new{(8) References Ref1: \cite{GP2024}, Ref2: \cite{BUBLAZAR2022}, Ref3: \cite{Park2024}, in which the corresponding images were first published; observations labelled `This work' are presented here for the first time.}
}
\label{tab:obs}
\end{table*}

\section{Methods and results \label{Sec:MethodsResults}}
\subsection{Total intensity images}
\begin{figure*}
\centering
\subfigure{\includegraphics[height=1.1\columnwidth]{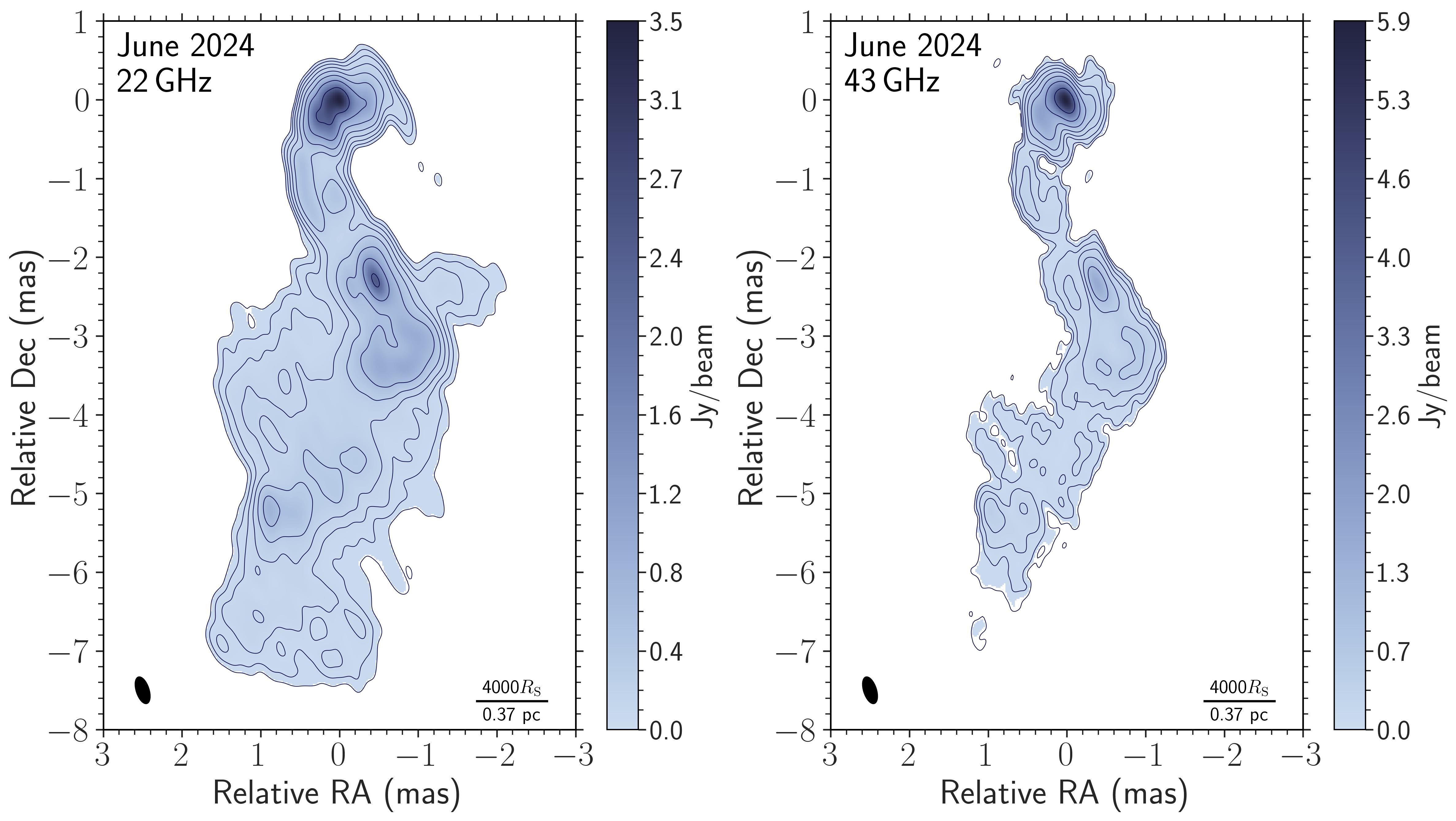}}
\caption{Total intensity images of \C\,at 22 and 43\,GHz. The total intensity is represented by the contours, using the contour levels at 0.5, 1, 2, 4, 8, 16, 32, and 64\% of the peak flux ($S_\mathrm{max, 22\,GHz}=3.54\,\mathrm{Jy/beam}$, $S_\mathrm{max, 43\,GHz}=5.94\,\mathrm{Jy/beam}$). The black ellipse in the bottom left corner denotes the common convolving beam with a size of $(0.35 \times 0.16)$\,mas at a position angle of $18^\circ$ (uniform weighting) and the black dash in the bottom right corner denotes the projected distance corresponding to $4000\,R_\mathrm{S}$ for both frequencies. 
The cut-off is at $4 \sigma_\textrm{I}$ for 22\,GHz and $1.5 \sigma_\textrm{I}$ for 43\,GHz ($\sigma_\textrm{I, 22\,GHz}=0.93\,\textrm{mJy},\,\sigma_\textrm{I, 43\,GHz}=4.9\,\textrm{mJy}$). 
}
\label{Fig:StokesIFig2024}
\end{figure*}

The total flux at 22\,GHz was scaled up to 90\% of the total flux density measured of single dish observations with the Effelsberg telescope~\citep[for a detailed procedure see][]{GP2024}.
When comparing the peak fluxes in Stokes I between the EVN and BEAM-ME data for 2024, we noted discrepancies, which can result from differences in amplitude calibration procedures, instrumental, or observational conditions. 
To correct for these offsets, we determined a scaling factor by comparing the peak fluxes of compact, unresolved features common in both data sets.
Specifically, we scaled the EVN fluxes at 43\,GHz by a factor of 2.0, which was calculated based on the comparison of the peak fluxes between the EVN and BU data observed on June 8, 2024.  
This scaling factor was then applied to the respective data set to ensure consistent flux measurements, allowing for a reliable and quantitative comparison of the jet structure and spectral properties. 
The total intensity images of 2021 and 2022 are already published by \cite{GP2024} and \cite{Park2024}, respectively. 
The Stokes I images of \C\, observed by the EVN in 2024 at 22\,GHz and 43\,GHz are shown in Fig.~\ref{Fig:StokesIFig2024}. 

\subsection{Spectral index distribution\label{subsec:SpecResults}}
\begin{figure*}
    \centering
    \includegraphics[width=1\textwidth]{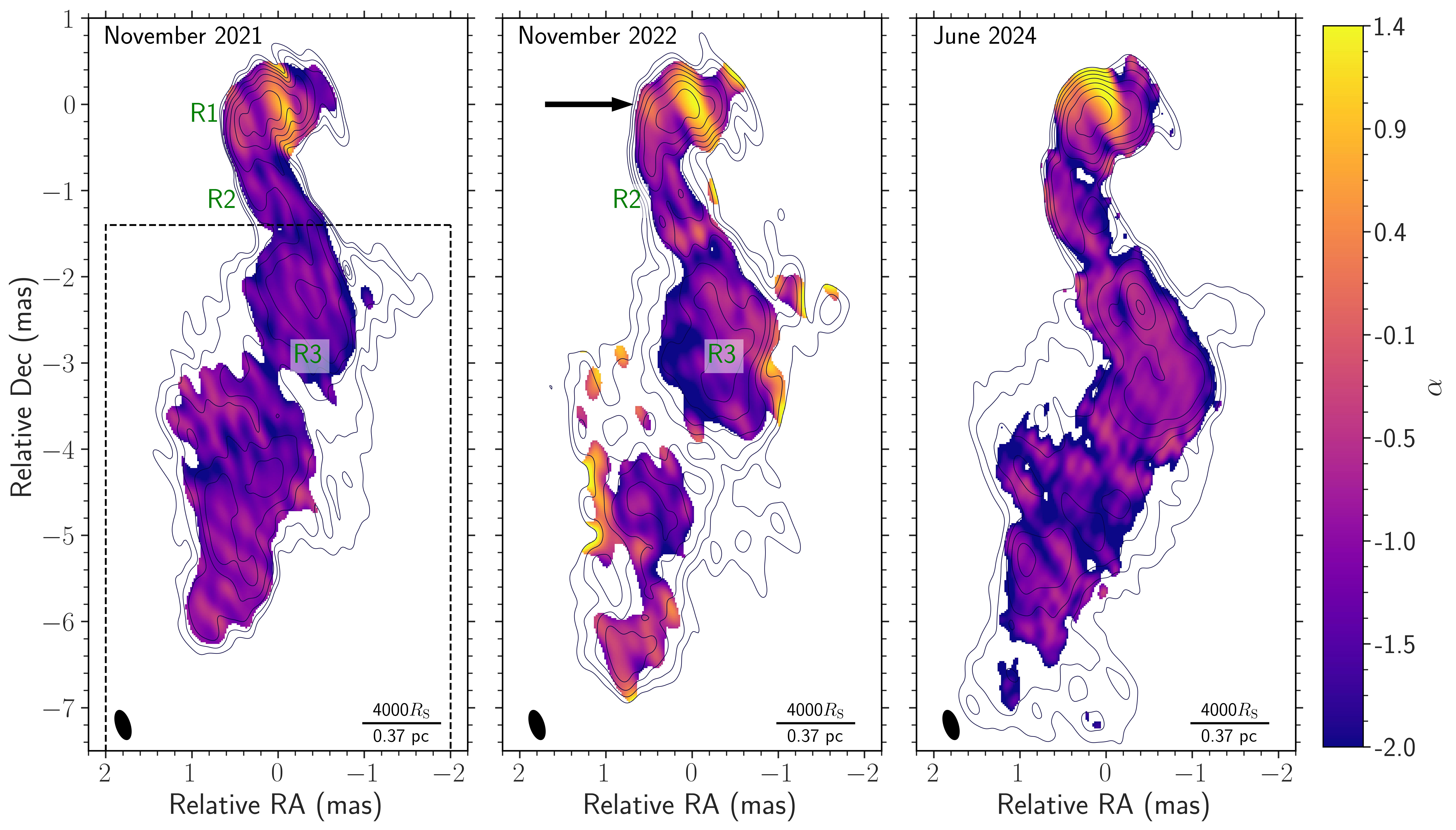}
    \caption{Spectral maps of \C\, between 22 and 43\,GHz. The total intensity is represented by the contours, using the contour levels at 1, 2, 4, 8, 16, 32, and 64\% of the peak flux at 22\,GHz, with a cut-off of $8 \sigma_\textrm{I}$ for all epochs. 
    The black ellipse in the bottom left corner denotes the convolving, circular beam size of $(0.35 \times 0.16)$\,mas at a position angle of $18^\circ$ for all epochs and the black dash in the bottom right corner denotes the projected distance corresponding to $4000\,R_\mathrm{S}$ and 0.37\,pc. 
    The dashed box indicates the optically thin jet region, which was used for the cross-correlation calculations as described in Sect.~\ref{subsec:SpecResults}. 
    The black arrow marks the knot deflection point as mentioned in Sect.~\ref{subsec:SpecResults}
    }
    \label{fig:Specmaps}
\end{figure*}

We define the spectral index $\alpha$ as $S_\nu \propto \nu^{+\alpha}$, with $S_\nu$ being the flux density and $\nu$ the observing frequency. 
In Fig.~\ref{fig:Specmaps} we present the results of the spectral analysis of 2021, 2022, and 2024, which were generated from total intensity images at 22 and 43\,GHz, convolved with the same beam size of $(0.35 \times 0.16)$\,mas at a position angle of $18^\circ$. 
After the phase self-calibration procedure of VLBI data, the absolute position information is lost and images at different frequencies need to be aligned manually, which is done using prominent and optically thin parts within the map~\citep{Kutkin2013}. 
We followed the procedure as presented in~\cite{GP2021} and performed a 2D cross-correlation on the optically thin part of \C's jet~\citep{Croke2008, Nagai2014, Park2024}, which is marked by the black dashed box in the first epoch in Fig.~\ref{fig:Specmaps}, and applied the shift in right ascension and declination to the lower frequency. 
For 2021 we find the best alignment of $0.06$\,mas and $-0.02$\,mas in right ascension and declination, respectively. 
Our alignment for 2022, with shifts of 0.02\,mas in right ascension and $0.0$\,mas in declination, is in agreement with the shift reported by~\cite{Park2024}. 
For 2024 we found the best alignment in right ascension and declination of $0.06$\,mas and $-0.06$\,mas, respectively. 
To evaluate the robustness of the alignment, we varied the sizes and positions for the alignment region and tested changes in alignments using a larger circular beam (see \ref{app:SpecmapsCIRC}), concluding that our results remain unaffected. 
We note that each pixel in the spectral index distribution map has an uncertainty of 10\%.
\\
In all three epochs we observe an inverted and flat spectrum in the core region of \C. 
Following the jet downstream, the spectral index gradually decreases, which was also observed in previous studies~\citep{GP2021, GP2022, Park2024}. 
In 2021 and 2022, the eastern edge of the first $\sim0.5~\mathrm{mas}$ of the inner jet region (region R1 and region marked with an arrow in Fig.~\ref{fig:Specmaps}) downstream of the jet is flat and inverted. 
In~\cite{Park2024} this region in 2022 is denoted as `the knot deflection point', coinciding with the region where a jet component changes drastically its direction~\citep[for more information see][]{Park2024}. 
Furthermore, in the first two epochs within the first $\sim1.5~\mathrm{mas}$ downstream from the core, we observe flattened spectra that extend as two limbs along the jet direction (regions R2 in Fig.~\ref{fig:Specmaps}). 
\new{This morphology corresponds to a limb‑brightened spectral index distribution, where the jet limbs exhibit flatter or inverted spectra, while the spectrum towards the confined jet edges is steeper. }
In 2024, the spectral index in the core region of \C\, is flat, while the downstream jet in 2024 is steep and shows an inverted spectrum \new{only in the east limb} in the innermost jet or pc-scale region. 
\new{To evaluate the spectral index evolution in the region of the limbs, we determine the spectral indices along slices at $\sim1.2~\mathrm{mas}$, and along parallel-shifted slices at $\sim1.0~\mathrm{mas}$ and $\sim1.4~\mathrm{mas}$ jet downstream, which are indicated by the black dashed lines in Fig.~\ref{app:SpecmapsSlices}. }
In the first two epochs, the limb structure is evident in the inset plot, while it is absent \new{for the west limb} in the last epoch. 
Therefore, we can conclude that the above mentioned limb-structure observed in 2021 and 2022 is less pronounced in 2024. 

\subsection{Light-curve information}
\begin{figure}
    \centering
    \includegraphics[width=0.45\textwidth]{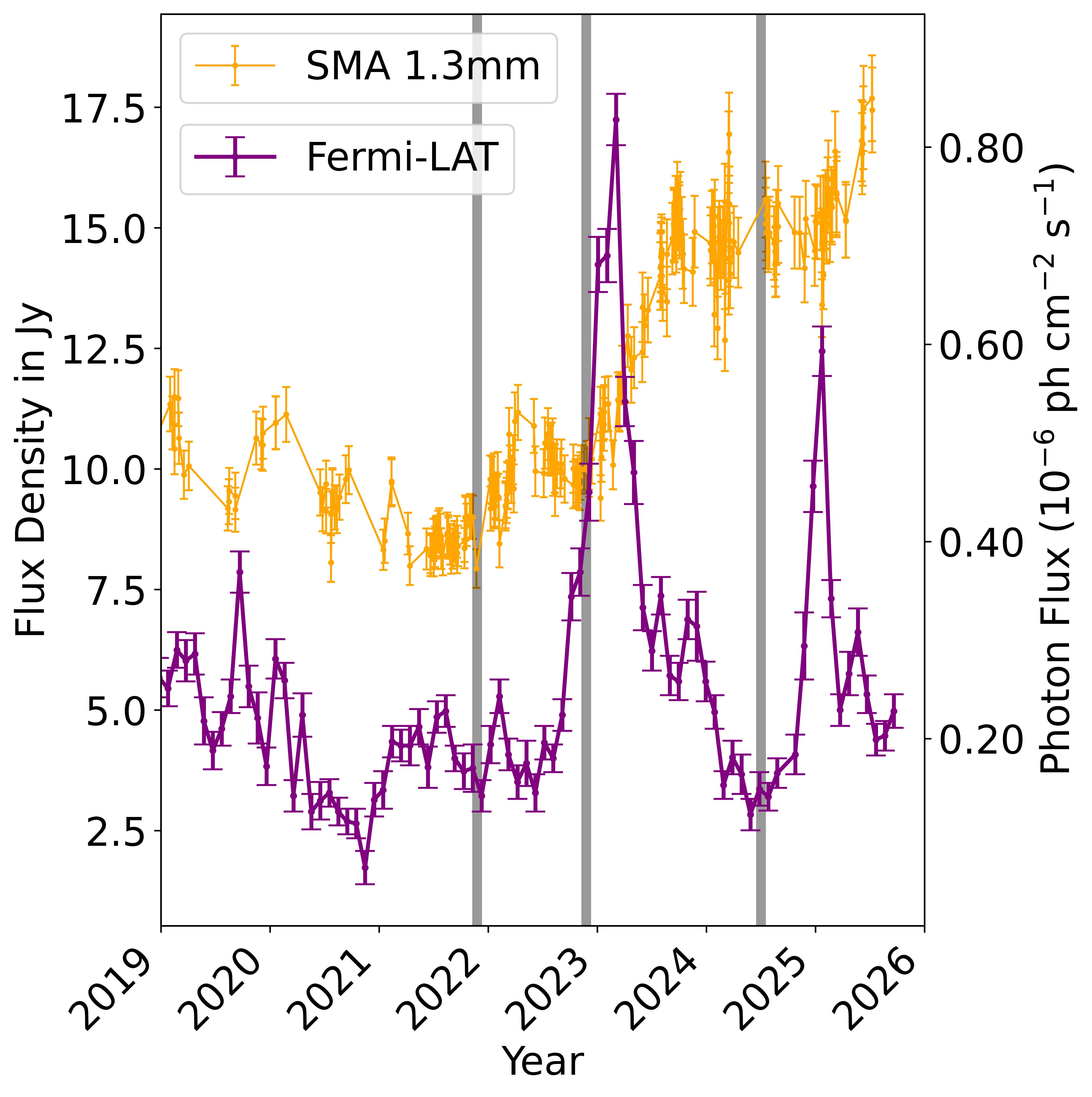}
    \caption{Light-curves of \C\,between 2019 and 2025. The orange curve denotes data by SMA at 1.3\,mm, the purple denotes monthly binned Fermi-LAT data and the grey shaded vertical lines showcase the epochs of available observational VLBI data at 22 and 43\,GHz. 
    }
    \label{fig:lightcurve}
\end{figure}

We utilised the $\gamma$-ray light-curve of \C\, in order to investigate whether morphological and spectral changes are associated with the emission of high energy photons. 
In Fig.~\ref{fig:lightcurve} we present the radio and $\gamma$-ray light-curves of \C\, employing publicly available data by SMA and Fermi-LAT between 2019 and 2025. 
Between 2019 and mid 2022, both radio and $\gamma$-ray light-curves do not show signs of flaring events. 
After 2023, the flux density in the radio regime starts to increase by a factor of $\sim1.5$ until the end of the available data. 
For the Fermi-LAT data, however, we observe a flare starting in June 2022, until it reaches its maximum in March 2023 with a peak of $(0.83\pm0.03)\times10^{-6}~\mathrm{ph~ cm^{-2}s^{-1}}$, after which the photon flux decreases. 
A more detailed examination of how this high-energy variability relates to filamentary structure and spectral index changes in the jet is given in Sect.~\ref{Sec:Discussion}.

\subsection{Magnetic field strength}
To estimate the magnetic field strength we assumed a conical jet with an optically thin spectral index $\alpha_0$, and equipartition between the radiating particle and the magnetic field, leading to a core shift factor of $k_\mathrm{r}=1$~\citep{Lobanov1998spectral, Hirotani2005, Ricci2022}. 
We can calculate the core-position offset measure as in~\cite{Hirotani2005}:
\begin{equation}
\Omega_{r\nu} = 4.85 \times 10^{-9} \,
\frac{\Delta r_{\mathrm{mas}} \, D_\mathrm{L}}
{(1 + z)^2}
\left(
\frac{\nu_1^{1/k_\mathrm{r}} \, \nu_2^{1/k_\mathrm{r}}}
{\nu_2^{1/k_\mathrm{r}} - \nu_1^{1/k_\mathrm{r}}}
\right)
\quad \mathrm{[pc\,GHz]},
\end{equation}
with $\Delta r_{\mathrm{mas}}$ being the core shift between frequency $\nu_1$ and $\nu_2$ in mas ($\nu_1 < \nu_2$), and $D_\mathrm{L}$ is the luminosity distance in pc. 
We found core-position offset measures of $\Omega_\mathrm{r\nu}^{2021} = 2.36\pm0.12~\mathrm{pc\,GHz}$, $\Omega_\mathrm{r\nu}^{2022} = 0.33\pm0.02~\mathrm{pc\,GHz}$, and $\Omega_\mathrm{r\nu}^{2024} = 1.39\pm0.07~\mathrm{pc\,GHz}$. 
Further, we calculated the magnetic field in the core region of \C\, following \cite{Fromm2013}:
\begin{align}
B_0 \approx &\frac{2 \pi m_\mathrm{e}^2 c^4}{e^3}
\left[\frac{e^2}{m_\mathrm{e} c^3} 
\left( \frac{\Omega_\mathrm{r\nu}}{r_0 \sin \theta} \right)^{k_\mathrm{r}}\right]^{\frac{5 - 2\alpha_0}{7 - 2\alpha_0}}
\Biggl[\pi C(\alpha_0) \frac{r_0 m_\mathrm{e} c^2}{e^2} \frac{-2\alpha_0}{\gamma_{\min}^{2\alpha_0 + 1}}\\
&\times \frac{\phi}{\sin \theta}K(\gamma, \alpha_0) 
\left(\frac{\delta}{1 + z}\right)^{\frac{3}{2} - \alpha_0}\Biggr]^{\frac{-2}{7 - 2\alpha_0}}\quad \mathrm{[G]},\nonumber
\end{align}
in which 
\begin{equation}
K(\gamma, \alpha_0) \sim
\frac{2\alpha_0 + 1}{2\alpha_0}
\frac{ \left(\gamma_{\max}/\gamma_{\min}\right)^{2\alpha_0} - 1 }
{ \left(\gamma_{\max}/\gamma_{\min}\right)^{2\alpha_0 + 1} - 1 }. \label{eq:K}
\end{equation}
The coefficient $C(\alpha_0)$ is tabulated in~\cite{Hirotani2005}, $r_0$ is the distance from the VLBI core to the jet apex and corresponds to the core-shift ($r_0^{2021}=0.14\pm0.01\,\mathrm{mas}$, $r_0^{2022}=0.02\pm0.001\,\mathrm{mas}$, $r_0^{2024}=0.08\pm0.01\,\mathrm{mas}$), $\gamma_{\max}$ and $\gamma_{\min}$ are the maximum and minimum Lorentz factor, $\phi$ is half the opening angle, $\theta$ the jet viewing angle, and $\delta=[\Gamma(1-\beta\cos\theta)]^{-1}$ is the Doppler factor. 
We adapted the parameters reported by~\cite{GP2021} $\delta\sim 1.18-1.25$, $\theta\sim 20^\circ-65^\circ$, $\phi\sim 2.8^\circ-20^\circ$, and $\gamma_{\max}=10^3-10^5$~\citep{Abdo2009} and $\gamma_{\min}=1$. 
For the limit $\alpha_0 \to -0.5$, the right hand side of $K(\gamma, \alpha_0)$ in Eq.~\ref{eq:K} converges to $1/\mathrm{ln}[\gamma_{\max}/\gamma_{\min}]$~\citep{Hirotani2005}. 
Finally, the magnetic field in 2021 is $B_0^{2021} \approx (30.6-75.8)~\mathrm{G}$, in 2022 $B_0^{2022} \approx (50.2-124.2)~\mathrm{G}$, and in 2024 $B_0^{2024} \approx (35.0-86.6)~\mathrm{G}$. \\
Furthermore, we can calculate the magnetic field strength in the region of the filaments at $\sim 1.2\,\mathrm{mas}$ (corresponding to 0.44\,pc) downstream from the core. 
At this distance, the jet remains optically thick, meaning that the relation above is still valid. 
We applied
\begin{equation}
    B=B_0(r_0/r_{1.2\,\mathrm{mas}})^b,
\end{equation}
where $r_{1.2\,\mathrm{mas}}$ is the distance of $\sim 1.2\,\mathrm{mas}$ from the core and $b=1$. 
For 2021 we computed $B_{1.2\,\mathrm{mas}}^{2021} \approx (3.7-9.1)~\mathrm{G}$, in 2022 $B_{1.2\,\mathrm{mas}}^{2022} \approx (0.8-2.1)~\mathrm{G}$, and in 2024 $B_{1.2\,\mathrm{mas}}^{2024} \approx (2.5-6.1)~\mathrm{G}$. 
We found that the magnetic field strength in both the core region and in the limbs is consistent within their ranges over three years.

\section{Discussion\label{Sec:Discussion}}
\new{The multi-epoch VLBI observations presented here reveal novel morphological and spectral details of the temporal evolution of limb-brightened jet structures close to the core of \C\, over three years, providing new constraints of its launching and collimation processes. }
We find a prominent inverted spectrum in the core region in 2021 and 2022, with a gradual steepening further downstream the jet. 
In these epochs, the inverted spectrum on the east edge of the inner jet region coincides with the knot deflection point, suggesting free-free absorption due to a dense cold ambient medium~\citep[for a detailed description see][]{Kino2021, Park2024}. 
We detect a limb-brightened morphology within the first $\sim1.5~\mathrm{mas}$ jet downstream, which is also imprinted in the spectral index distribution as flattened spectra in 2021 and 2022 (R2 in Fig.~\ref{fig:Specmaps}). 
\new{Another bright radio galaxy with a similar limb-brightened geometry is M\,87, where intertwined helical threads align with a flattened spectral index~\citep{Nikonov2023}, which can be caused by K-H instabilities~\citep{Lobanov2003, Hardee2003}. }
The observed flattened spectra in region R2 and R3 in Fig.~\ref{fig:Specmaps} coincide with \new{the areas of possible overlapping filaments} as seen in~\cite{GP2025}, where the authors observe brightness enhancements at $\sim$1\,mas and $\sim$3\,mas from the core. 
We interpret these overlaps as a plausible explanation for the observed morphology. 
\new{They further suggest, that the jet shows an increased emissivity in R3, which could be due to the local decrease of the separation of the two filaments, as an alternative approach to the overlapping filaments. }
\\
In 2024, however, the \new{west limb of the double rail} structure in the downstream jet is less pronounced~(see Fig.~\ref{app:SpecmapsSlices}).  
A possible explanation for this behaviour is the rotation of filaments in the innermost jet region as discussed in \cite{GP2025} and as seen in the \href{https://www.aanda.org/articles/aa/olm/2024/05/aa47562-23/aa47562-23.html}{animated sequence} by~\cite{Park2024}: 
In 2021 and 2022, these filaments might overlap, inducing a brightness enhancement in the regions where these filaments cross, which is then imprinted in the spectral index distribution as flattened spectra. 
In 2023, however, these overlapping filaments can be the cause of a $\gamma$-ray flare as can be seen from Fig.~\ref{fig:lightcurve}. 
\new{Based on the analysis presented in \citeauthor{Hodgson2021}~(\citeyear{Hodgson2021}; and references therein), this behaviour can be interpreted within a turbulence induced magnetic reconnection scenario, in which filamentary magnetic structures produced behind a travelling shock fronts interact and reconnect. 
As the disturbance (R3) propagates downstream, increasing turbulence amplifies and disturbs the magnetic field as filaments, enhancing the probability of reconnection events and the formation of `mini‑jets' \citep[e.g.][]{Giannios2009, Shukla2020}. 
Such reconnection events can efficiently accelerate particles and produce high‑energy emission on sub‑parsec scales, potentially within the limb region of the jet \citep{Hodgson2021,Giannios2013}. 
We therefore suggest that the $\gamma$ rays may be produced in the limbs spanning from the core to $\sim2~\mathrm{mas}$ downstream. } 
As these filaments continue to rotate, the positions of the overlap shift and might not induce brightness enhancements \new{in the latest epoch.} 
In addition, synchrotron cooling can lead to a steepening of the spectrum downstream from the core~\citep{Nikonov2023}. \\
\new{The observed limb-brightening in the spectral index distribution across all epochs suggests a transverse stratification of the inner jet, consistent with a magnetically dominated flow close to the central engine. 
General relativistic magnetohydrodynamic (GRMHD) simulations show that this morphology is consistent with a fast-spinning BH with magnetic field lines reaching down to the central engine~\citep{Takahashi2018}. 
Moreover, the observed geometry can be associated with a magnetically arrested disc (MAD) and advection-dominated accretion flows~\citep[ADAFs; see][]{Narayan1995ADAF} in which spin energy is extracted from the central engine, favouring a Blandford–Znajek launching process~\citep[see][for more details]{Tchekhovskoy2011, GP2024EHT}. }
We want to note the limitations of this model, as a standard and normal evolution (SANE) model can produce similar geometries, such as a toroidal magnetic field configuration, but the limb-brightened structure is most prominent in simulations for the MAD model~\citep{Fromm2022}.
\\
\new{Recent general relativistic magnetohydrodynamic simulations in the parsec scale jet of M\,87 show that a fast-spinning black hole in a MAD state agrees well with the observed jet morphology~\citep{Yang2024}. }
Similarly, this provides further support to the interpretation that \C’s jet is also powered by the Blandford–Znajek mechanism. 
Further, we find that the magnetic field strengths overlap within their ranges over three epochs and do not show a high variability within three years. \\
\new{Previous studies of the magnetic field strength of the jet apex in \C\, by~\cite{GP2021, GP2022} reported lower values than the estimates presented in this work, differing by up to an order of magnitude, whereas the estimate by \cite{Kim2019} is of the same order of magnitude as our results.  
However, \cite{Nagai2017} estimated the magnetic field strength at the jet base of \C\, applying a first-order approximation of a radial magnetic field configuration ($B(r)\propto r^{-2}$) leading to $\sim10^4$\,G. }
\new{Similar estimates were conducted by \cite{OSullivan2009}, analysing multiple BL-Lac objects, leading to magnetic field strength estimates of $\sim10^4-10^5$\,G. 
Furthermore, \cite{Kino2015} report a magnetic field strength of $50-124$\,G at the jet base of M\,87, which is similar to the computed values presented in this work. 
The magnetic field strength variability across different studies can be explained by the fact that these estimates are sensitive to model parameters (i.e. $r_0$ and $\Omega_\mathrm{r\nu}$) and the method used to determine the field strength, and can therefore vary across multiple epochs. }\\
Although we adopt a conical jet for the calculation of the magnetic field strength, we want to emphasise the limitations of this model. 
The Stokes\,I maps in Fig.~\ref{Fig:StokesIFig2024} indicate that, on sub-parsec scales, the jet appears more confined and exhibits a morphology closer to a cylindrical shape than to a cone but expands to a wider, conical jet further downstream~\citep{Nagai2014, Giovannini2018, Foschi2025}. 
\new{Besides the conical jet model, \cite{Oh2022} also tested a parabolical jet model to estimate the black hole location in \C\, and concluded that the available data do not allow for a robust discrimination between the two models. 
Moreover, the jet geometry cannot be uniquely determined from a qualitative evaluation of Stokes\,I images alone, as projection effects, Doppler boosting, and transverse stratification can produce similar apparent morphologies. 
Given the different assumptions adopted in these studies and the uncertainties associated with the observational determination of the models, it is therefore not yet possible to draw a strong conclusion on the most probable jet geometry on sub‑parsec scales in \C. 
We thus adopt the conical jet model as the most physically consistent model and commonly used framework for \C\, and to enable direct comparison with previous VLBI studies of AGN jets \citep{Blandford1979, Oh2022, GP2022}.
}

\section{Conclusions \label{Sec:Conclusion}}
In this study, we present a detailed analysis of the innermost jet structure of \C\, using multi-frequency, multi-epoch VLBI observations at 22 and 43\,GHz. 
Our results provide new constraints on the jet’s spectral index properties, with particular attention to the limb-brightened features near the core region.
Our results can be summarised as follow:
\begin{itemize}
    \item We observe a significant change in the spectral index distribution over three epochs: in 2021 and 2022, the core region shows inverted spectra with limb-brightened, flattened spectral structures in the first few milliarcseconds jet downstream; in 2024 the core spectrum is inverted while the region up to $\sim1.5~\mathrm{mas}$ jet downstream exhibits a steeper spectral index \new{with the west limb of the previously observed limb‑brightened structure being significantly less pronounced.}
    \item We detect persistent limb-brightening in the sub-parsec region across all epochs, \new{suggesting that the observed geometry is} produced by a fast-spinning black hole in a MAD/ADAF state, favouring a Blandford–Znajek jet launching scenario. 
    \item \new{The limb-brightened and flat spectra in 2021 and 2022 can be explained by overlapping filaments}; the non-detection \new{of the west limb} in 2024 may be linked to rotational motion of those filaments, synchrotron cooling after a $\gamma$-ray flare, or changes in jet-ambient medium interactions. 
    \item The correlation between the spectral evolution, filament rotation, and the $\gamma$-ray light-curve supports a scenario in which the high-energy emission originates from the limbs.
    \item We measured the core-shifts leading to magnetic field strengths in the range of $(0.8-9.1)~\mathrm{G}$ in the region of the limbs. 
\end{itemize}
These results show that the inner jet of \C\ is a highly dynamic structure where spectral properties, morphology, and magnetic field strength vary on timescales of a few years, likely modulated by filamentary variations, and interactions between the jet and the ambient medium.

\begin{acknowledgements}
\new{We would like to thank the anonymous referee for their constructive comments, which improved our work.} 
We thank I. Liodakis for valuable comments and insightful discussions. 
\new{We would like to thank J. Park for providing the data of 2022 and for helpful information and comments on our study. }
The research leading to these results has received funding from the European Union’s Horizon 2020 research and innovation program under grant agreement No 101004719 [Opticon RadioNet Pilot ORP]. 
The European VLBI Network is a joint facility of independent European, African, Asian, and North American radio astronomy institutes. 
Scientific results from data presented in this publication are derived from the following
EVN project code: GP058, GP061.
This study makes use of VLBA data from the VLBA-BU Blazar Monitoring Program (BEAM-ME and VLBA-BU-BLAZAR; \href{http://www.bu.edu/blazars/BEAM-ME.html}{http://www.bu.edu/blazars/BEAM-ME.html}), funded by NASA through the Fermi Guest Investigator Program. 
J.~A.~H. acknowledges the support of the National Research Foundation of Korea (NRF) (NRF-2021R1C1C1009973) and that this work was supported by the National Research Foundation of Korea (NRF) grant funded by the Korea government (MSIT) RS-2025-16302968. 
M.~M.~L was supported by the FONDECYT Iniciac\'on grant 11251078. 
J.~Y.~K. is supported for this research by the National Research Foundation of Korea (NRF) grant funded by the Korean government (Ministry of Science and ICT; grant no. 2022R1C1C1005255, RS-2022-NR071771) and by the Korea Astronomy and Space Science Institute under the R\&D program (Project No. 2025-9-844-00) supervised by the Korea AeroSpace Administration. 
The Submillimeter Array is a joint project between the Smithsonian Astrophysical Observatory and the Academia Sinica Institute of Astronomy and Astrophysics and is funded by the Smithsonian Institution and the Academia Sinica. 
We recognize that Maunakea is a culturally important site for the indigenous Hawaiian people; we are privileged to study the cosmos from its summit. 
This research is based in part on observations obtained with the 100-m telescope of the MPIfR at Effelsberg, observations carried out at the IRAM 30-m telescope operated by IRAM, which is supported by INSU/CNRS (France), MPG (Germany) and IGN (Spain), observations obtained with the Yebes 40-m radio telescope at the Yebes Observatory, which is operated by the Spanish Geographic Institute (IGN, Ministerio de Transportes, Movilidad y Agenda Urbana), and observations supported by the Green Bank Observatory, which is a main facility funded by the NSF operated by the Associated Universities. We acknowledge support from the Onsala Space Observatory national infrastructure for providing facilities and observational support. The Onsala Space Observatory receives funding from the Swedish Research Council through grant no. 2017-00648. This publication makes use of data obtained at the Metsähovi Radio Observatory, operated by Aalto University. 
This publication acknowledges project M2FINDERS, which is funded by the European Research Council (ERC) under the European Union’s Horizon 2020 research and innovation programme (grant agreement no. 101018682).  
\end{acknowledgements}
\bibliographystyle{aa} 
\bibliography{aanda} 
\begin{appendix}
\section{Additional information}
\begin{figure}[h]
    \begin{minipage}{\textwidth}
        \centering
        \includegraphics[width=\textwidth]{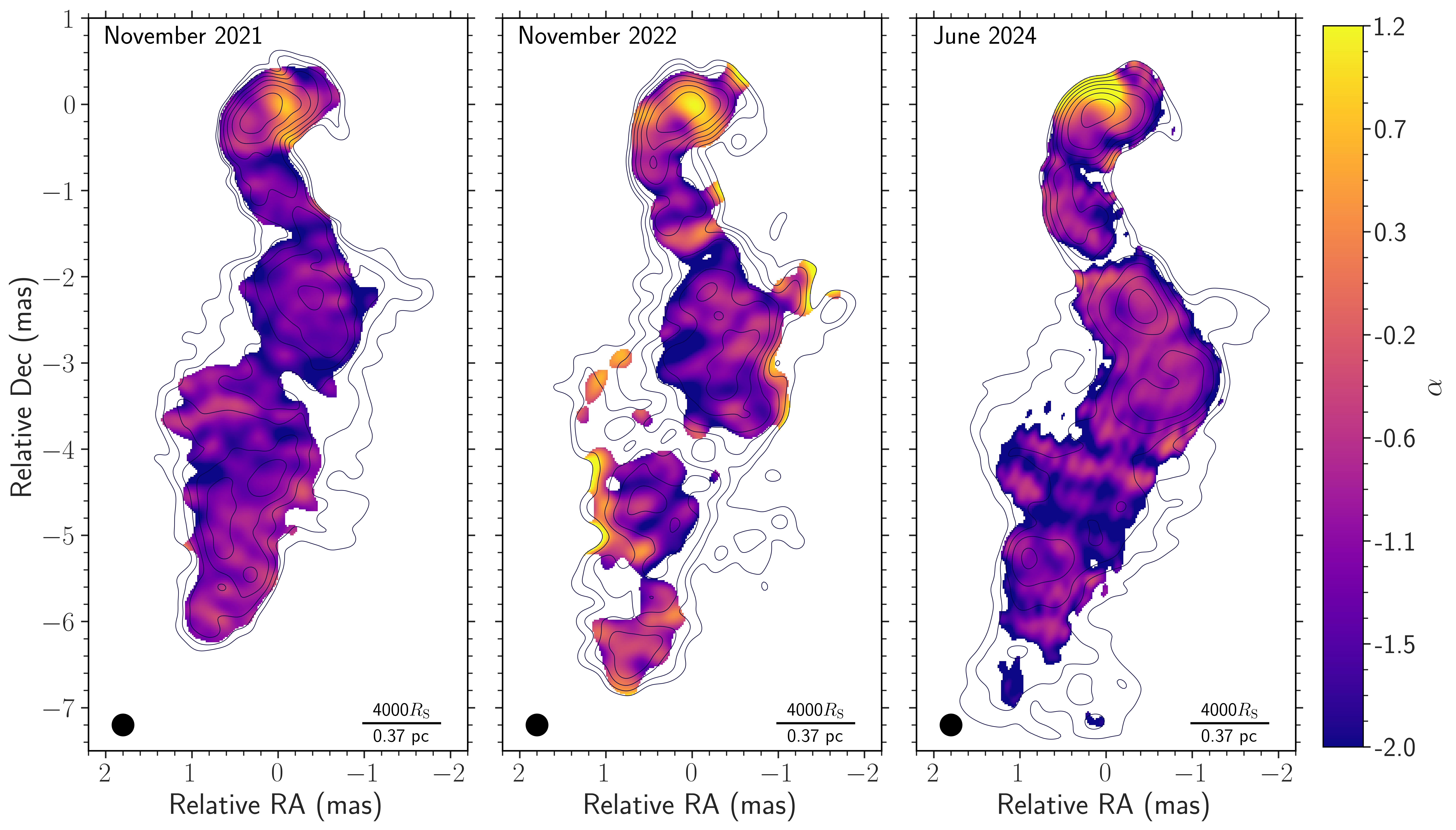}
        \caption{Spectral maps of \C\, between 22 and 43\,GHz. The total intensity is represented by the contours, using the contour levels at 1, 2, 4, 8, 16, 32, and 64\% of the peak flux at 22\,GHz, with a cut-off of $8 \sigma_\textrm{I}$ for all epochs. 
        The black circle in the bottom left corner denotes the convolving, circular beam size of 0.25\,mas for all epochs and the black dash in the bottom right corner denotes the projected distance corresponding to $4000\,R_\mathrm{S}$ and 0.37\,pc. 
        }
        \label{app:SpecmapsCIRC}
    \end{minipage}
    \end{figure}

\begin{figure}[htb]
    \begin{minipage}{\textwidth}
        \centering
        \includegraphics[width=\textwidth]{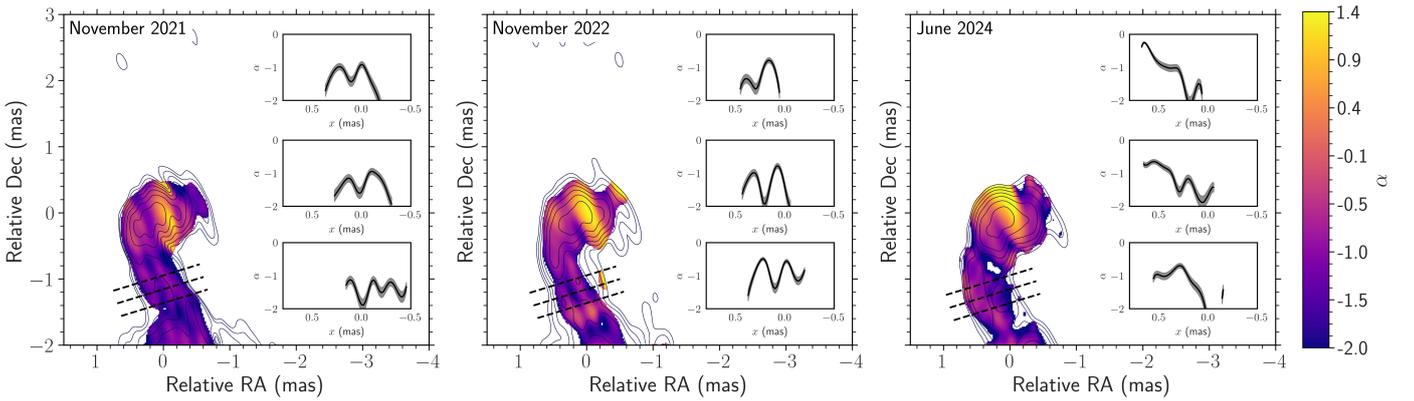}
        \caption{\new{Similar to Fig.~\ref{fig:Specmaps}, but the inset figures show the spectral indices distributions along slices at $\sim1.2~\mathrm{mas}$, and along parallel-shifted slices at $\sim1.0~\mathrm{mas}$ and $\sim1.4~\mathrm{mas}$ jet downstream, as indicated by the black dashed lines. The grey shaded area around the inset plot indicates the 10\% uncertainty of the spectral index. }}
        \label{app:SpecmapsSlices}
    \end{minipage}
    \end{figure}

Figure~\ref{app:SpecmapsCIRC} presents the same images as in Fig.~\ref{fig:Specmaps} but with a circular convolving beam of 0.25\,mas and Fig.~\ref{app:SpecmapsSlices} shows the spectral index distribution along \new{slices at $\sim1.2~\mathrm{mas}$, and along parallel-shifted slices at $\sim1.0~\mathrm{mas}$ and $\sim1.4~\mathrm{mas}$ jet downstream.} 
The discussion of these map are covered in the main text. 

\end{appendix}
\end{document}